\documentstyle[12pt]{article}

\begin{document}
{\it University of Shizuoka}

\hspace*{10cm} {\bf US-93-08}\\[-.3in]

\hspace*{10cm} {\bf September, 1993}\\[.4in]

\begin{center}

{\large\bf Phenomenological Quark Mass Matrix Model }\\[.1in]
{\large\bf with Two Adjustable Parameters}\\[.5in]

{\bf Yoshio Koide} \\[.1in]

Department of Physics, University of Shizuoka \\[.1in]
395 Yada, Shizuoka 422, Japan \\[.5in]

{\large\bf Abstract}\\[.1in]

\end{center}

\begin{quotation}
A phenomenological quark mass matrix model which includes only two
adjustable parameters is proposed from the point of view of the
unification of quark and lepton mass matrices.
The model can provide reasonable values of  quark mass ratios
and Kobayashi-Maskawa matrix parameters.
\end{quotation}
\newpage

It is widely accepted that the family number of ordinary quarks and leptons
is three.
(This does not ruled out a possibility that there are some extraordinary
families, e.g. a family with an extremely heavy neutrino, and so on.)
Then, we have ten observable quantities related to up- and down-quark
mass matrices, $M_u$ and $M_d$, i.e.,
six up- and down-quark masses and four parameters of Kobayashi-Maskawa
(KM) [1] matrix.
On the other hand, most of quark mass matrix models currently proposed
include adjustable parameters more than five (two parameters for
each quark mass matrix $M_q$ ($q=u,d$)  and one relative phase
parameter between up- and down-quark mass matrix phase parameters).
At present, every model is comparably plausible, and is in agreement with
the present experimental data.
Nevertheless, we cannot resist the temptation to investigate a further
new-type mass matrix form of $(M_u,M_d)$ with parameters less than four,
because we expect that the quark and lepton families are governed by
a more fundamental law of the nature.

In the present paper, we propose the following model of
quark and lepton mass matrices inspired by an extended technicolor-like
model:
$$ M_f = m_0 \, G O_f G \ ,  \eqno(1) $$
$$ G = {\rm diag}(g_1, g_2, g_3) \ , \eqno(2) $$
$$ O_f = {\bf 1}+ 3a_f X(\phi_f) \ , \eqno(3) $$
$$
{\bf 1}= \left(
\begin{array}{ccc}
1 & 0 & 0 \\
0 & 1 & 0 \\
0 & 0 & 1
\end{array} \right) \ , \ \ \ \ \
X(\phi)=\frac{1}{3} \, \left(
\begin{array}{ccc}
1 & e^{i\phi} & 1 \\
e^{-i\phi} & 1 & 1 \\
1 & 1 & 1
\end{array} \right) \ , \eqno(4)
$$
where $f=\nu, e, u$, and $d$ are indices for neutrinos, charged leptons,
up- and down-quarks, respectively.
Here, the diagonal matrix $G$ denotes a coupling constant matrix of
a hypercolored boson $\phi_\alpha$ with ordinary fermions $f_i$ and
hypercolored fermions $F_{i\alpha}$ ($\alpha$ and $i$ are
hypercolor and family indices, respectively), and the matrix $O_f$
denotes the condensation of the hypercolored fermions
$\langle (\overline{F}F)\rangle$.
Since we consider the so-called
seesaw mechanism [2] for neutrino mass matrix,
the matrix $M_\nu$ given in (1) should be taken
as the Dirac mass matrix part of the neutrino mass matrix.

As we discuss below, since we take $a_e=0$ in the charged lepton mass
matrix $M_e$,
the parameters $g_i$ are fixed as $\sqrt{m_0}G=$diag$(\sqrt{m_e},
\sqrt{m_\mu},\sqrt{m_\tau})$, so that
the mass matrix $M_f$ is
effectively given by
$$
M_f= \left(
\begin{array}{ccc}
m_e & 0 & 0 \\
0 & m_\mu & 0 \\
0 & 0 & m_\tau
\end{array} \right) + a_f \left(
\begin{array}{ccc}
m_e & e^{i\phi_f}\sqrt{m_e m_\mu} & \sqrt{m_e m_\tau} \\
e^{-i\phi_f}\sqrt{m_e m_\mu} & m_\mu & \sqrt{m_\mu m_\tau} \\
\sqrt{m_e m_\tau} & \sqrt{m_\mu m_\tau} & m_\tau
\end{array} \right) \ . \eqno(5)
$$
In the present paper, we put an ansatz for $\phi_f$,
$(\phi_u=0,\ \phi_d=\pi/2)$, so that
adjustable parameters in the quark mass matrices
$M_u$ and $M_d$ are  only two, $a_u$ and $a_d$.
As we demonstrate later, a suitable choice of the parameters  $a_u$ and $a_d$
will provide not only reasonable values of up- and down-quark mass ratios
$m^u_i/m^u_j$  and $m^d_i/m^d_j$ $(i,j=1,2,3)$, respectively, but also
reasonable values of the ratios $m^u_i/m^d_i$
as well as reasonable values of KM matrix parameters.

The mass matrix forms $M_e$ and $M_\nu$ have already proposed by
the author [3,4] from the phenomenological point of view.
In fact, the present quark mass matrix model was inspired by the
phenomenological success of the charged and neutrino mass matrices
as we review below.

Ten years ago, the author [3] has proposed a charged lepton mass
matrix model, in which charged lepton masses $m^e_i=(m_e,m_\mu,m_\tau)$ are
generated through the condensations of hypercolored fermions $E_{i\alpha}$,
$\langle (\overline{E}E)\rangle$,
and the exchanges of a hypercolored vector boson $\phi_\alpha$ which is
coupled with $\sum_i g_i\overline{e}_i E_{i\alpha}$, i.e.,
the masses $m^e_i$ are given by
$m^e_i\simeq g_i^2 \langle (\overline{E}E)\rangle /m_\phi^2$.
(The model is similar to the extended technicolor model [5], but we
consider that the vector boson $\phi_\alpha$ is not a gauge boson.)
Here, the hypercolored boson $\phi_\alpha$ (hereafter we drop the index
$\alpha$) is a particularly mixed state among $SU(3)$-family octet bosons
$\phi_3$ and $\phi_8$ and singlet boson $\phi_0$, which are the
$\lambda_3$, $\lambda_8$ and $\lambda_0$ components of $SU(3)$.
We consider that the octet bosons acquire large masses at an energy scale
$\Lambda_H$ except for one component $\phi^{(8)}$ which is a liner
combination of $\phi_3$ and $\phi_8$, while  $\phi^{(8)}$
has exactly the same mass as $\phi_0$.
Then, if there is a mixing term between $\phi^{(8)}$ and $\phi_0$,
the 45$^\circ$ mixing between $\phi^{(8)}$ and $\phi_0$ is inevitably
caused.
We assume that only one of the two states can contribute to
the mass matrix $M_e$, so that the coupling constant $g_i$ is
given by $g_i=(g_i^{(8)}+g_0)/\sqrt{2}$, where
$g_i^{(8)}$ and $g_0$ are coupling constants of $\phi^{(8)}$
and $\phi_0$ with $\overline{e}_i E_{i}$, respectively, and
they satisfy the relations $g_1^{(8)}+g_2^{(8)}+g_3^{(8)}=0$
and $(g_1^{(8)})^2+(g_2^{(8)})^2+(g_3^{(8)})^2=3g_0^2$, because we
consider that $\phi^{(8)}$ and $\phi_0$ belong to the nonet of $U(3)$-family.
Then, the coupling constants $g_i$ satisfy the relation
$$ g_1^2+g_2^2+g_3^2=\sum_{i=1}^3\left( \frac{g_i^{(8)}+g_0}{\sqrt{2}}
 \right)^2=3 g_0^2=\frac{2}{3}\left( \sum_{i=1}^3
\frac{g_i^{(8)}+g_0}{\sqrt{2}}\right)^2=\frac{2}{3}(g_1+g_2+g_3)^2 \ ,
\eqno(6) $$
which leads to a charged lepton mass sum rule [3]
$$
m_e+m_\mu+m_\tau=\frac{2}{3}(\sqrt{m_e}+\sqrt{m_\mu}+\sqrt{m_\tau})^2
\ . \eqno(7)
$$
The sum rule (7) predicts $m_\tau=1776.97$ MeV
from the input values of $m_e$ and $m_\mu$.
The predicted value 1777 MeV is in excellent agreement with
the observed values of $m_\tau$ which have recently been reported
by ARGUS [6], BES [7] and CLEO [8] collaborations.
Thus, the phenomenological success of the charged lepton mass matrix
$M_e = m_0\, G {\bf 1} G $ is our main motivation to consider the mass
matrix form of the type  $m_0 GO_fG$.

In Ref.~[3], the boson state $\phi$ is more explicitly given by
$$
\phi=-\frac{1}{\sqrt{2}} \left[ \cos(\frac{\pi}{4}-\epsilon)\, \phi_3
- \sin(\frac{\pi}{4}-\epsilon)\, \phi_8 \right]
+\frac{1}{\sqrt{2}} \, \phi_0 \ . \eqno(8)
$$
As a result, the matrix $G$ is given by
$$
G =  \frac{1+\varepsilon}
{2\sqrt{2}\sqrt{1+\varepsilon^2}} \, \left(
\begin{array}{ccc}
0 & 0 & 0 \\
0 & -1 & 0 \\
0 & 0 & 1
\end{array} \right) +
\frac{1-\varepsilon}{2\sqrt{6}\sqrt{1+\varepsilon^2}} \, \left(
\begin{array}{ccc}
-2 & 0 & 0 \\
0 & 1 & 0 \\
0 & 0 & 1
\end{array} \right) $$
$$  +
\frac{1}{\sqrt{6}} \, \left(
\begin{array}{ccc}
1 & 0 & 0 \\
0 & 1 & 0 \\
0 & 0 & 1
\end{array} \right)  \ , \eqno(9)
$$
where  $\cos(\pi/4-\epsilon)$ and $\sin(\pi/4-\epsilon)$ are replaced by
$(1+\varepsilon)/\sqrt{2(1+\varepsilon^2)}$ and
$(1-\varepsilon)/\sqrt{2(1+\varepsilon^2)}$, respectively.
In the limit of ``ideal mixing", i.e., $\varepsilon=0$,
the model leads to massless electron.
This explains why electron mass is extremely small
compared with other charged lepton masses.

The motivation to consider the matrix form $O_f$ of the type
${\bf 1}+3a_fX$ is as follows:
Recently, in order to explain a neutrino mixing value
$\sin\theta_{e\mu}\simeq 0.04$
($\sin^22\theta_{e\mu}\simeq 7 \times 10^{-3}$) suggested by
GALLEX [9], the author [4] has proposed a neutrino mass matrix model,
in which the neutrino mass matrix $M_\nu$ is given by
$M_\nu\simeq M_\nu^D M_M^{-1}M_\nu^D=(M_\nu^D)^2/m_M$
($m_M$ is a Majorana neutrino mass) on the basis of
the conventional seesaw mechanism scenario [2], and
the Dirac mass matrix $M_\nu^D$ is given by the form
$M_\nu^D = m^\nu_0\, G ({\bf 1}+ 3 a_\nu X(0)) G$,
where $a_\nu$ is a numerical parameter with $a_\nu\gg 1$.
Here we have supposed that the hypercolored neutrino condensation
$\langle (\overline{N}N)\rangle$ takes
a democratic term ($X(0)$-term) dominance form, ${\bf 1}+3a_\nu X(0)$,
differently from the case of
$\langle (\overline{E}E)\rangle \propto {\bf 1}$.
The model can lead to a desirable prediction [4]
$ \sin\theta_{e\mu}\simeq (1/2)\sqrt{m_e/m_\mu}\simeq 0.035$
for $a_\nu \gg 1$.

In general, the mass matrix (5) with $\phi_f=0$ provides
the relation [4]
$${m^f_1}/{m^f_2} \simeq {3m_e}/{4m_\mu}=0.00363 \ , \eqno(10)$$
in the limit of $1/a_f\rightarrow 0$,
where $m_i$ are eigenvalues of the mass matrix (5) and are
defined as $|m_1|< |m_2|< |m_3|$.
The conventional values [10] of the running quark masses at 1 GeV,
$m_u\simeq 5.1$ MeV and $m_c\simeq 1.35$
GeV, provide $m_u/m_c\simeq 0.0038$, which is in agreement with
the prediction (10).
This is just the motivation to consider the mass matrix $M_u$ given by
(1) with $\phi_u=0$, i.e.,
$$
M_u=m_0^u G\left( {\bf 1}+3 a_u X(0)\right) G \ . \eqno(11)
$$

Hereafter, for convenience, we will refer
the Gasser--Leutwyler's values [10]
for $\Lambda_{\overline{MS}}^{(3)}=0.150$ GeV [11] as running quark mass
values (in unit of GeV) at an energy scale 1 GeV:
$$
\begin{array}{lll}
m_u=0.0051 \pm 0.0015 \ , & m_c=1.35 \pm 0.05 \ , & m_t=226^{+43}_{-49} \ , \\
m_d=0.0089 \pm 0.0026 \ , & m_s=0.175 \pm 0.055 \ , & m_b=5.58 \pm 0.13 \ . \\
\end{array} \eqno(12)
$$
The value of $m_t(1$ GeV), which is not listed in the original paper
by Gasser and Leutwyler, has been estimated by using
the standard model parameter fitting value $m_t^{phys}=130^{+25}_{-28}$ GeV
[12] and $\Lambda_{\overline{MS}}^{(3)}=0.150$ GeV
($\Lambda_{\overline{MS}}^{(4)}=0.114$ GeV,
$\Lambda_{\overline{MS}}^{(5)}=0.0699$ GeV).
However, the values in (12) should not be taken rigidly, because
the estimates are highly dependent on the value of $\Lambda_{\overline{MS}}$
and models (prescriptions) at present [13].

The mass matrix given by (11) actually can predict reasonable
up-quark mass ratios: for example, $m_u/m_c=0.00389$ (0.00379) and
$m_c/m_t=0.00597$ ($-0.00598$) for $a_u=16.45$ ($a_u=-19.02$).
Here, since the quark mass ratio $m_u/m_c$ is insensitive
to the value of $a_u$, we have determined the value of
$a_u$ from the value of $m_c/m_t$ in (12).
The prediction of $m_u/m_c$ is in excellent agreement with the
value of $m_u/m_c$ provided by (12).

Next, we seek for a mass matrix form for down-quarks.
We cannot choose the same mass matrix form as that for up-quarks,
i.e., $M_d=m^d_0\, G({\bf 1}+3 a_d X(0))G$,
because it leads to a wrong down-quark
mass ratio $m^d_1/m^d_2\simeq 3m_e/4m_\mu$.
Besides, we must introduce a $CP$ violation phase to the model.
We assume a down-quark matrix form $M_d$ which is similar to $M_u$,
but which has a phase factor $\phi_d\neq 0$ as given by (4).

In general, the eigenvalues $m_i^f$ of the mass matrix (5) are given by
$$
\frac{m_1^f}{m_\tau}\simeq \frac{\kappa_f(3+\kappa_f)-4\sin^2(\phi_f/2)}{
\kappa_f^2(2+\kappa_f)}\, \varepsilon_1 \ , \ \
\frac{m_2^f}{m_\tau}\simeq \frac{2+\kappa_f}{1+\kappa_f}\, \varepsilon_2
\ , \ \ \frac{m_3^f}{m_\tau}\simeq \frac{1+\kappa_f}{\kappa_f} \ ,
\eqno(13)$$
where $\kappa_f=1/a_f$, $\varepsilon_1=m_e/m_\tau$ and
$\varepsilon_2=m_\mu/m_\tau$.
Note that, differently from the case of $M_u$ with $\phi_u=0$,
we cannot take a limit of $\kappa_d\rightarrow 0$ in the down-quark mass
matrix $M_d$, because the mass ratio $m_1^d/m_2^d$ includes a factor
$1/\kappa_d^2$.
The relation
$$
\frac{m_s}{m_b}\simeq \frac{(2+\kappa_d)\kappa_d}{1+\kappa_d}\,
\frac{m_\mu}{m_\tau} \eqno(14)
$$
suggests a small but visible value  $\kappa_d\simeq -0.2$
because $|m_s/m_b|\simeq 0.03$ and $m_\mu/m_\tau\simeq 0.06$.
Then, the relation
$$
\frac{m_d m_s}{m_b^2}\simeq - \frac{4}{(1+\kappa_d)^3}\,
\frac{m_e m_\mu}{m_\tau^2} \, \sin^2\frac{\phi_d}{2}  \eqno(15)
$$
suggests $|\phi_d|\simeq \pi/2$.
For simplicity, we fix $\phi_d$ to be $\phi_d=\pi/2$,
which leads to a maximal $CP$ violation.

In conclusion, we assume that the down-quark mass matrix $M_d$ is
given by
$$
M_d = m^d _0\, G \left( {\bf 1}+3 a_d X(\frac{\pi}{2})\right) G
\ . \eqno(16)
$$
Then, a suitable choice of $a_d$ can provide
excellent predictions of $m_d/m_s$ and $m_s/m_b$:
for example, $m_d/m_s=-0.0507$ and $m_s/m_b=-0.0313$ for $a_d=-4.81$.
It is noted that, in the mass matrix $M_f$ with $\phi_f=\pi/2$,
in general, two values of $a_f$, $a_f=a_f^{(1)}$ and
$a_f=a_f^{(2)}$, which satisfy the relation
$(1/a_f^{(1)})+(1/a_f^{(2)})=-2$, can yield the same mass ratios
$m^f_1/m^f_2$ and $m^f_2/m^f_3$.
Therefore, the alternative choice $a_d=a_d^{(2)}= -0.558$
provides the same predictions of the down-quark mass ratios
as the case of $a_d=a_d^{(1)}=-4.81$.

The quark mass matrix model $(M_u, M_d)$ given in (11) and (16)
predicts the KM matrix elements $V_{ij}$ in the limit of
$1\ll |\kappa_d| \ll |\kappa_u| \rightarrow 0$ as follows:
\setcounter{equation}{16}
\begin{eqnarray}
|V_{us}|^2 & \simeq & 2\frac{1+\kappa_d}{(2+\kappa_d)^2\kappa_d^2}\,
\frac{m_e}{m_\mu} \ ,  \\[.1in]
|V_{cb}|^2 & \simeq & \frac{\kappa_d^2}{(1+\kappa_d)^2}\, \frac{m_\mu}{m_\tau}
\simeq \frac{m_e/m_\tau}{|V_{us}|^2} \ ,  \\[.1in]
|V_{ub}|^2 & \simeq & \frac{m_e}{m_\tau} \ .
\end{eqnarray}
The relation (17) leads to the well-known Weinberg--Fritzsch empirical
relation [14] $|V_{us}|\simeq \sqrt{-m_d/m_s}$, because the mass ratio
$m_d/m_s$ is given by
$$
\frac{m_d}{m_s}\simeq -\frac{(1+\kappa_d)^2(2-2\kappa_d-\kappa_d^2)}{
(2+\kappa_d)^2\kappa_d^2}\, \frac{m_e}{m_\mu} \ . \eqno(20)
$$
The predicted values of $|V_{cb}|$ and $|V_{ub}|$ from (18) and (19) are
somewhat large compared with the observed values.
This disagreement comes from the approximation in which we took $\kappa_u=0$.
The values of $|V_{ij}|$ are sensitive to $\kappa_u$ and $\kappa_d$
as well as to $\varepsilon_1$ and $\varepsilon_2$.
As we demonstrate below, a suitable choice of $\kappa_u=1/a_u$
can predict reasonable values of $|V_{cb}|$ and
$|V_{ub}|$ numerically.

In Table I, we show predictions on the KM matrix parameters
for the values of $a_u$ and $a_d$ which provide reasonable quark mass
ratios.
We also list the prediction of the rephasing invariant quantity $J$ [15].
The case  $a_d=a_d^{(1)}= -4.81$ can provide
reasonable values of the KM matrix parameters
except that the value of $|V_{ub}|$ is somewhat small.
The value of $|V_{ub}|$ is highly sensitive to the value of the phase
parameter $\phi_d$ when $M_d$ is given by
$M_d=m^d_0\, G({\bf 1}+3a_d X(\phi_d))G$, and a choice of $\phi_d$ slightly
different from $\phi_d=\pi/2$ predicts a fairly large value of $|V_{ub}|$
compared with the case of exact $\phi_d=\pi/2$.
It is likely that the prediction of $|V_{ub}|$ becomes
reasonable value by renormalization effects for $M_q$.

On the other hand, the second case $a_d=a_d^{(2)}= -0.558$
cannot provide reasonable values of $|V_{cb}|$ and
$|V_{ub}|$ as seen in Table I.
However, it should be noted that the case  $a_d=-0.558$ can provide
not only the excellent predictions of $m_d/m_s$ and $m_s/m_b$ but also
the excellent prediction of $m_d/m_u$ if we consider $m_0^u=m_0^d$.
When we put $m^d_2=0.175$ GeV (i.e., $m_0^u/m_0^e=m_0^d/m_0^e=6.52$)
in order to compare our prediction
with the Gasser--Leutwyler's values (12),
we obtain the following quark mass
values at energy scale 1 GeV for the case of $(a_u, a_d)=(-19.02,-0.558)$:
$$
\begin{array}{lll}
m^u_1=0.00504 \ {\rm GeV} \ , & m^u_2=+1.33 \ {\rm GeV} \ , &
m^u_3=-223 \ {\rm GeV} \ , \\
m^d_1=0.00887 \ {\rm GeV} \ , & m^d_2=-0.175 \ {\rm GeV} \ , &
m^d_3=+5.59 \ {\rm GeV} \ .  \\
\end{array}   \eqno(21)
$$
The values (21) are in excellent agreement with the Gasser--Leutwyler's
values (12).
In most of the conventional quark mass matrix models,
if we want to explain the fact $m_t\gg m_b$,
then we must be contented with saying that the fact $m_u\sim m_d$ is
an accidental coincidence in the model.
In the case of $a_d=a_d^{(2)}$, we can obtain the reasonable ratio of
$m_u/m_d$ together with the reasonable ratios $m^u_i/m^u_j$ and
$m^d_i/m^d_j$.
Therefore, the case of $a_d=-0.558$ is worth being taken into consideration
as well as the case $a_d=-4.81$.

It should be also be noted that predictions of $|V_{ij}|$ in the
case of $a_d^{(2)}$ are, in general,  exactly the same as
those in the case of $a_d^{(1)}$ if we take
$$
      V= U_u P U_d^\dagger \ ,  \eqno(22)
$$
$$
    P={\rm diag}(1, 1, -1) \ , \eqno(23)
$$
instead of $V=U_u U_d^\dagger$.
The modification (22)  means that the mass matrices $(M_u; M_d)$
given by (11) and (16) are not
those for the weak eigenstate quark basis $(u_0, c_0, t_0; d_0, s_0, b_0)$,
but those for the quark basis $(u_0, c_0, \pm t_0; d_0, s_0, \mp b_0)$.
Although the origin of such phase inversion is not clear,
if we accept the scenario, we can provide not only the reasonable
values (21) of quark masses but also reasonable values of the
KM matrix parameters
$$
|V_{us}|=0.203 \ , \ \ \ |V_{cb}|=0.0393 \ , \ \ \ |V_{ub}|=0.00139 \ ,
\ \ \ |V_{td}|=0.00882 \ , $$
$$ J=0.891 \times 10^{-5} \ , \eqno(24)
$$
by fixing $(a_u, a_d)=(-19.02, -0.558)$.

So far, we have neglected the energy scale dependence of
the quark masses and KM parameters.
We consider that the mass matrix form (1) is given at
an energy scale $\mu=M_X$.
We expect that fine tuning of our parameters in consideration of
the renormalization group equations can provide further excellent
predictions of quark masses and KM mixing parameters.

We consider that $m_0^q$ and $m_0^e$ satisfy $m_0^u=m_0^d=m_0^e$
at the energy scale $M_X$ and
the value $(m_0^q/m_0^e)_{\rm 1 GeV}=6.52$ will be explained
by evolving $m_0^q$ and $m_0^e$ down from $M_X$ to 1 GeV.
In the present model, the energy scale $M_X$ need not
be identical with the weak boson mass scale $v\simeq 250$ GeV.
In order to give rough estimate of $M_X$, we neglect electroweak
interaction and use, for convenience,  the equation for
QCD running quark mass (for example, see Ref.~[10])
(not the renormalization group equation for the Yukawa couplings).
The value of $M_X$ estimated is highly sensitive to the choice of
$\Lambda_{QCD}$ ($\Lambda^{(n)}_{\overline{MS}}$).
If we adopt a recent experimental value
$\Lambda^{(4)}_{\overline{MS}}=0.260$ GeV [16]
($\Lambda^{(3)}_{\overline{MS}}=0.311$ GeV,
$\Lambda^{(5)}_{\overline{MS}}=0.175$ GeV,
$\Lambda^{(6)}_{\overline{MS}}=0.0709$ GeV), we obtain
$M_X \sim 10^{18}$ GeV.
The value of $M_X$ is somewhat large.
However, the present estimate of $M_X$ is only a trial
and it should not be taken seriously.
The estimate is also highly dependent on the models.
In order to give more accurate estimate of $M_X$,
we must build the model more concretely.

In conclusion, we have proposed a phenomenological quark and lepton mass
matrix model (1).
The matrix form (1) has a possibility of unified description of
quark and lepton masses and their mixings.
The mass matrix form $m_0GO_fG$ can be understood from an extended
technicolor-like scenario (but our boson $\phi_\alpha$ is not a gauge
boson).
However, such a mass matrix form (1) can also be understood from
a Higgs-boson scenario with some additional $U(1)$ charges.
In both scenarios, it is essential that there are heavy fermions which
behave as intermediate states in the mass generation mechanism of the
light fermions.
In the derivation of the sum rule (7), it is essential that
the 45$^\circ$ mixing between octet and singlet parts in the
$U(3)$-family nonet scheme.
In Ref.~[17], the sum rule (7) has been re-derived from
a Higgs potential model with a mixing term between $SU(3)$-family octet
and singlet.
However, Ref.~[17] did not discuss clearly on the additional $U(1)$
charges which should be introduced in the scenario.
Recently, a detailed study of the $U(1)$ charges related to
the horizontal symmetry has been given by Leurer, Nir and Seiberg [18].
We will find a clue to the justification of the present scenario
in their paper, in which we can see relations of $|V_{ij}|$
similar to our relations (17)--(19), although in our model the parameters
$\kappa_u$ and $\kappa_d$ are not negligible.
However, the purpose of the present paper is to propose
a new-type mass matrix form (1), and not to give a reasonable
mass generation mechanism for the mass matrix form (1).
Theoretical justification of the model (1)
will be given elsewhere.

\vglue.3in

\centerline{\bf Acknowledgments}
The author would like to thank Prof.~M.~Tanimoto for careful reading
the manuscript and helpful comments.
This work has been supported in part by a Shizuoka Prefectural
Government Grant.

\vglue.3in
\newcounter{0000}
\centerline{\bf References and Footnotes}
\begin{list}
{~\arabic{0000}~.}{\usecounter{0000}
\labelwidth=1cm\labelsep=.4cm\setlength{\leftmargin=1.7cm}
{\rightmargin=.4cm}}
\item M.~Kobayashi and T.~Maskawa, Prog.~Theor.~Phys. {\bf 49}, 652 (1973).
\item M.~Gell-Mann, P.~Rammond and R.~Slansky, in {\it Supergravity},
eds. P.~van Nieuwenhuizen and D.~Z.~Freedman (North-Holland, 1979);
T.~Yanagida, in {\it Proc.~Workshop of the Unified Theory and Baryon
Number in the Universe}, eds. A.~Sawada and A.~Sugamoto (KEK, 1979).
\item Y.~Koide, Phys.~Rev. {\bf D28}, 252 (1983).
\item Y.~Koide, Mod.~Phys.~Lett. {\bf 8}, 2071 (1993).
\item S.~Dimopulous and L.~Susskind, Nucl.~Phys. {\bf B155}, 237 (1979);
E.~Eichten and K.~D.~Lane, Phys.~Lett. {\bf B90}, 125 (1980).
\item  ARGUS collaboration, H.~Albrecht {\it et al}.,
Phys.~Lett. {\bf B292}, 221 (1992).
\item  BES collaboration, J.~Z.~Bai {\it et al}., Phys.~Rev.Lett. {\bf 69},
3021 (1992).
\item CLEO collaboration, R.~Balest {\it et al}., Phys.~Rev. {\bf D47},
R3671 (1993).
\item GALLEX collaboration, P.~Anselman {\it et al}., Phys.~Lett. {\bf B285},
390 (1992).
\item J.~Gasser and H.~Leutwyler, Phys.~Rep. {\bf 87}, 77 (1983).
\item Usually the value $m_b=5.3\pm 0.1$ GeV is referred as the
Gasser--Leutwyler's value of $m_b$.  However, the value is not
that for $\Lambda_{\overline{MS}}^{(3)}=0.150$ GeV.
\item J.~Ellis, G.~L.~Fogli and E.~Lisi, Phys.~Lett. {\bf B274}, 456 (1992).
Also see, Z.~Hioki, Mod.~Phys.~Lett. {\bf 6}, 2129 (1991).
\item For a recent study of light quark masses, for example, see,
J.~F.~Donoghue and B.~E.~Holstein, Phys.~Rev.~Lett. {\bf 69}, 3444 (1992).
\item S.~Weinberg, Ann.~N.~Y.~Acad.~Sci. {\bf 38}, 185 (1977);
H.~Fritzsch, Phys.~Lett. {\bf 73B}, 317; Nucl.~Phys. {\bf B155}, 189 (1979).
\item C.~Jarlskog, Phys.~Rev.~Lett. {\bf 55}, 1839 (1985);
O.~W.~Greenberg, Phys.~Rev. {\bf D32}, 1841 (1985);
I.~Dunietz, O.~W.~Greenberg, and D.-d.~Wu, Phys.~Rev.~Lett. {\bf 55},
2935 (1985);
C.~Hamzaoui and A.~Barroso, Phys.~Lett. {\bf 154B}, 202 (1985);
D.-d.~Wu, Phys.~Rev. {\bf D33}, 860 (1986).
\item Particle Data Group, K.~Hikasa {\it et el}., Phys.~Rev. {\bf D45},
S1 (1992).
\item Y.~Koide, Mod.~Phys.~Lett. {\bf A5}, 2319 (1990).
\item M.~Leurer, Y.~Nir and N.~Seiberg, Nucl.~Phys. {\bf B398}, 319 (1993).
\end{list}
\vglue.3in
\vfill

Table I. \ Prediction on the KM matrix parameters.
$$
\begin{array}{cccccccc}\hline\hline
a_u & a_d & |V_{us}|  &  |V_{cb}| & |V_{ub}|  &  |V_{td}|
&  |V_{ub}/V_{cb}| & J \\ \hline
+16.45 & -4.81 & 0.204  & 0.0624  & 0.00186  & 0.01291 & 0.0298
& 2.31\times 10^{-5}  \\
-19.02 & -4.81 & 0.203  & 0.0393  & 0.00139  & 0.00882
& 0.0353 & 0.891 \times 10^{-5}
\\ \hline
+16.45 & -0.558 & 0.201  & 0.495  & 0.0170  & 0.0907  & 0.0344
& 1.10 \times 10^{-3} \\
-19.02 & -0.558 & 0.199  & 0.515  & 0.0169  & 0.0942  & 0.0328
& 1.12 \times 10^{-3} \\ \hline\hline
\end{array}
$$


\end{document}